\newcommand{\sr}{R_{\odot}}

\newcommand{\rsh}{{\partial \Omega/\partial r}}
\documentclass[oldversion]{aa}
\usepackage{graphicx}
\usepackage{txfonts}
\usepackage{natbib}
\bibpunct{(}{)}{;}{a}{}{,}

\begin{document}
   \title{How does the shape and thickness of the tachocline
    affect the distribution of the toroidal magnetic
    fields in the solar dynamo?}

   \subtitle{}

   \author{G. Guerrero
          \inst{1}
          \and
          E. M. de Gouveia Dal Pino\inst{1}\fnmsep\thanks{Just to show the usage
          of the elements in the author field}
          }

   \offprints{G. Guerrero}

   \institute{Astronomy Department, Instituto de Astronomia,
     Geof\'{i}sica e Ci\^{e}ncias Atmosf\^{e}ricas\\
     Universidade de S\~{a}o Paulo,
     Rua do Mat\~{a}o 1226, S\~{a}o Paulo, Brazil\\
     \email{guerrero,dalpino@astro.iag.usp.br}
   }

   \date{}


   \abstract
       {Flux-dominated solar dynamo models, which have
     demonstrated to be quite successful in reproducing most of the
     observed features of the large scale solar magnetic cycle,
     generally produce an inappropriate latitudinal distribution of the
     toroidal magnetic fields, showing fields of large magnitude in
     polar regions where the radial shear has a maximum
     amplitude. Employing a kinematic solar dynamo model, we here explore
     the contribution of both the radial and the latitudinal shear in
     the generation of the toroidal magnetic fields by varying the shape
     and the thickness of the solar tachocline. We also explore the
       effects of the diffusivity profile of the convective zone.
     Considering the shear term of the dynamo
     equation, $({\bf B_p} \cdot \nabla) \Omega$$=$$B_r\rsh +
     B_{\theta}/r \, \partial \Omega / \partial \theta$, we find that
     the latitudinal component is always dominant over the radial
     component at producing toroidal field amplification. 
These results are very sensitive to the adopted diffusivity
profile, specially in the inner convection zone (which is
caracterized by the diffusivity $\eta_c$ and the radius $r_c$ of
transition between a weak and a strong turbulent region). A
diagram of the toroidal field at a latitude of $60^{\circ}$ versus
the diffusivity at the convection layer for different values of
the tachocline width has revealed that these fields are mainly
eliminated  for tachoclines with width $d_1$$\gtrsim$$0.08\sr$
(for $\eta_c$$\simeq$$2\times10^9$ $-$ $1\times10^{10}$ cm$^2$
s$^{-1}$ and $r_c$$=$$0.715\sr$); or for $d_1$$\lesssim$$0.02\sr$ and
almost any value of $\eta_c$ in the appropriate solar range. For
intermediate values of $d_1$$\simeq$$0.04\sr$$-$$0.06\sr$, strong toroidal
fields should survive at high latitudes in the butterfly diagram
and those values are therefore not suitable. We have built
butterfly diagrams for both a thin and a
      thick tachocline that best match the observations.
We have also found that a prolate tachocline is able to reproduce
       solar-like butterfly diagrams depending on the choice of
       appropriate diffusivity profiles and tachocline width range.
}


   \maketitle


\section{Introduction}
Over the last few years, with the increase of the computational
power and the improvement in the observational techniques, there
has been a substantial advance in the knowledge of our magnetic
star. However there is still a considerable number of open
questions regarding the magneto-hydrodynamic processes that govern
the large scale solar magnetic phenomena, i.e., the 11-years
sunspot cycle, the polarity inversions of the magnetic field, the
phase lag between the toroidal and poloidal inversions, and the
latitudinal distribution of the toroidal magnetic field. In the
currently accepted scenario, the large scale solar magnetic cycle
(LSSMC) is governed by a dynamo action that is the responsible for
the transformation of a positively oriented poloidal magnetic
field in a negatively oriented toroidal field, and the subsequent
transformation of this one in another poloidal field but with the
opposed polarity, and so on, until completing the cycle. The first
stage of the process is the well known $\Omega$ effect: the
poloidal field lines are dragged (``{\it frozen in}''), and
amplified by the differential rotation of the fluid. From early
results of the helioseismology, there has been a common agreement
that this process takes place in the tachocline, a thin layer
($\le 5\%$ of the solar radii) located at the base of the solar
convection zone ($\sim 0.7 \sr$) where the transition from a
uniformly rotating regime of the radiative core to a
differentially rotating regime in the convective envelope occurs.
The strong radial shear ($\rsh$) that exists in this region
suggests that $\Omega$ acts mainly in the tachocline instead of in
the
entire convection zone.\\
The second stage of the process, conventionally called $\alpha$
effect from the early days of the Parker turbulent mean field
dynamo, consists in to convert this belt of toroidal field in a
new poloidal field with opposite polarity. This subject has been
the center of intense discussions and debates and an excellent
revision of both, the $\alpha$ effect mechanisms and the several
dynamo models can be found in  \cite{charbonneau2005} and
references there in. In this work we assume an $\alpha$ effect as
the result of the decay of active bipolar magnetic regions (BMR),
as it was originally proposed by \cite{babcock}  and followed by
\cite{leighton}. The fundamental ingredient in this mechanism is
the emergence across the convective bulk of magnetic flux tubes
due to the Parker-Rayleigh-Taylor instability in order to form
sunspots or active BMRs at the surface, while the BMRs decay they
migrate to the poles carrying with them the necessary magnetic
flux to neutralize the remnant poloidal field and generate a new
one. Magnetic reconnection between big loops of both hemispheres
happens in this phase, as observed.

Magnetic flux tube simulations
\citep{dsilva93,fanetal93,caligetal95,caligetal98,fanfisher96,fan04}
have shown that magnetic fields between $5\times10^4$ and $10^5$ G
are able to erupt across the convection zone and emerge to the
surface with the observed inclinations in the sunspots (joy's
law). A sub-adiabatic layer is necessary in order to store
magnetic field up to this strength and once again the
tachocline is the best place to allow this storage and
amplification. The decay of the BMRs at the surface requires both,
super-granular diffusion and meridional transport, for this reason
the numerical models of the solar dynamo that work in the
kinematic regime, that is, ignoring dynamical back-reaction of the
magnetic field on the flux, include four fundamental ingredients:
differential rotation, meridional circulation, diffusion terms and
a source of poloidal field (the $\alpha$ term).

Recent kinematic models in  the above scenario have been able to
reproduce the majority of the LSSMC features, but have failed at
producing a correct latitudinal distribution of the toroidal
field, showing intense magnetic fluxes at higher latitudes and,
consequently, sunspots close to the poles.  \cite{nandy2002} found
a possible solution to this problem by allowing the meridional
flow to penetrate below the tachocline. Under this hypothesis, the
magnetic flux will be stored in a highly sub-adiabatic region and
will emerge only at the desired latitudes. This assumption has
given rise to a new controversy: how much of the meridional flow
must penetrate below the tacholine? On one side, in a recent work
\cite{chatterjeeetal04}, working under this hypothesis, have been
able to reproduce the characteristics of the solar cycle in the
two hemispheres, and also the observed parity rule (Hale's law).
On the other side, numerical simulations of a meridional flow
penetrating in a sub-adiabatic medium \citep{gilmanmiesh04} have
shown that the dynamical effects of the fluid alone, without
including magnetic field, do not allow a penetration below $5\%$
of the tachocline and, in the case that the flow can penetrate in
some way the radiative zone, the penetration would be reduced to
about few kilometers only. In a recent report,
\cite{rudigeretal05} confirm this result arguing further that a
meridional circulation confined within the convection zone alone
would be able to produce an $\alpha \Omega$ dynamo. Another
problem that arises when the penetration of the flow in the
radiative zone is allowed is the excessive burning of light
elements in such a hot zone. Numerical models of mixing, based on
helioseismic measurements of the sound speed and density profiles,
indicate a maximum extent of mixing of $5\%$ inside the tachocline
\citep{brunetal2002}. Also, \cite{guerrero04} developed a hybrid
model using the profile of meridional velocities of
\cite{nandy2002} and the $\alpha$-term of \cite{dikchar99} and
found that a deep meridional flow solves only partially the
problem of the distribution of the toroidal fields at the polar
regions. Recently, \cite{dikpatietal04} have found an appropriate
combination of parameters that are able to better reproduce the
observations, however, until now there is no clear physical
process that is able to explain why the sunspots appear only at
lower latitudes.

In this work, we use a modified version of the numerical code
developed by \cite{guerrero04} introducing more detailed
diffusivity and $\alpha$ profiles, as suggested by
\cite{dikpatietal04}, aiming to explore how the shape and
thickness of the tachocline may influence the latitudinal
distribution of the toroidal field. Although during the entire
dynamo process there are several mechanisms that are still not
understood, such as, the exact values and the variations of the
magnetic diffusivity in the convective and radiative envelopes,
and the return flow properties, the actual radial and latitudinal
operation of the $\alpha$-effect, or the quantity of poloidal
field that can be produced by the decay of the BMRs, for our
propose here, we will consider profiles that are constrained by
helioseismology observations and also by new results of
simulations.

In section \S2 we present the basic mathematical formalism of the
solar dynamo mechanism and some details about the construction of
the model (a complete description of the numerical tool is given
in \cite{guerrero04}); in \S3, we present a detailed discussion of
the assumed profiles and the set of parameters that allow for a
best fit to the observations. The results of our analysis are
described in \S4; and finally in \S5 we sketch our conclusions.


\section{The model}

The equation that describes the spatial and temporal evolution of the
magnetic field in an ionized medium is the magnetic
induction equation, which under the assumption of azimuthal symmetry
can be divided in its poloidal and toroidal components, respectively:

\begin{eqnarray}
\frac{\partial A}{\partial t}+\frac{1}{s}({\bf u}
\cdot\nabla)(sA)=\eta\left(\nabla^2-\frac{1}{s^2}\right)A +
S_1(r,\theta,B)\label{eq1} \quad ,\\
\frac{\partial B}{\partial t}+\frac{1}{r}\left[\frac{\partial}{\partial
    r}(ru_rB)+\frac{\partial}{\partial
    \theta}(u_{\theta}B)\right]\label{eq2}=({\bf B_p} \cdot
\nabla) \Omega\\\nonumber
-\nabla \eta \times (\nabla \times
B)+\eta\left(\nabla^2-\frac{1}{s^2}\right)B\nonumber \quad,
\end{eqnarray}

\noindent where ${\bf B_p}$$=$$B_r+B_{\theta}$$=$$\nabla
\times A$ is the poloidal field and $B$ the
toroidal field, $\Omega$ is the angular velocity, ${\bf
  u}$$=$$u_r+ u_{\theta}$ denotes the
meridional components of the velocity field,
$s$$=$$r\sin\theta$, $\eta$ is the magnetic diffusivity,  and
$S_1(r,\theta,B)$ is a source that describes the
$\alpha$-effect in the Babcock-Leighton dynamo.

The above equations are solved in a 2-dimensional mesh of $128$
$\times$ $128$ grid points by using the ADI implicit method within
the physical domain: $0.55 \sr \le r \le \sr$, and $0^{\circ}$
(pole) $\le \theta \le 90^{\circ}$ (equator). The boundary
conditions are as follows: $B$ and $A$ are set to zero in the pole
and in the bottom radial boundary. At the top boundary we set a
vacuum condition, so that, $B(\sr,\theta)=0$ and $A(\sr,\theta)$
is smoothly matched with a vacuum photosphere-corona,
$\left(\nabla^2 - \frac{1}{r^2 \sin^2
  \theta}\right)A$$=$$0$. At the equator, we set the toroidal component,
$B$$=$$0$, and a continuous
boundary, $\frac{\partial A}{\partial  \theta}$$=$$0$, for the
poloidal field.


\section{The dynamo ingredients}

As discussed above, a kinematic solar dynamo model needs four
fundamental ingredients: differential rotation, meridional
circulation, magnetic diffusivity and a poloidal source term. The
characteristics of these ingredients are in part constrained by
helioseismology observations but there is some degree of freedom in
the choice of the parameters that may reproduce better the LSSMC
features.  We assume the following profiles and parameters in order to
generate a fiducial model able to reproduce the observed butterfly
diagram.

\subsection{Differential rotation}

One of the goals of the helioseismology was the determination of
an accurate profile for the angular velocity both at the surface
and in deeper layers. The results revealed a radiative core that
is rotating with uniform velocity. This changes to a convective
bulk that is rotating differentially with a retrograde velocity
with respect to the radiative interior at higher latitudes and a
pro-grade velocity at lower latitudes. The interface between these
two regimes is a thin layer called tachocline. An analytical
profile of this differential rotation was introduced by
\cite{charbonneaumcgregor97} and used for the first time in a
Babcock-Leighton type dynamo by \cite{dikchar99}. We employ here
the same expression used by \cite{dikchar99}:

\begin{equation}
\Omega(r,\theta)=\Omega_c+\frac{1}{2}\left[1+\mathrm{erf}
  \left(2\frac{r-R_c}{d_1}\right)\right]
(\Omega_s(\theta)-\Omega_c) \quad ,
\label{eq3}
\end{equation}

\noindent where $\Omega_c/2 \pi$$=$$432.8$ nHz is the uniform angular
velocity of the radiative core,
$\Omega_s(\theta)$$=$$\Omega_{eq}+a_2\cos^2  \theta+a_4\cos^4 \theta$
is the latitudinal differential rotation at
the surface with $\Omega_{eq}/2\pi$$=$$460.7$ nHz being the angular
velocity at the equator,  $a_2/2\pi$$=$$-62.9$ nHz and
$a_4/2\pi$$=$$-67.13$ nHz, erf$(x)$ is an error function that confines the
radial shear to a tachocline located in $R_c$$=$$0.7\sr$ of thickness
$d_1$. Figure 1 depicts the
differential rotation profile in the solar interior for the core and
convective layers.

   \begin{figure}
   \centering
   \includegraphics[width=6cm]{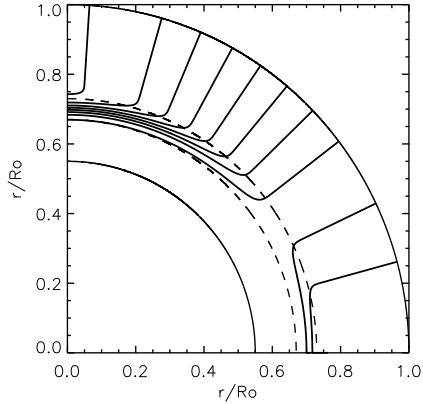}
   \caption{Isorotation lines of the solar interior inferred from
     helioseismological observations. The dashed lines
     show the approximate thickness of the tachocline.}
   \label{fig1}
   \end{figure}

Until now there has been no consensus about the location and
thickness of the tachocline \citep{cobardetal01}. Some
observational works \citep{antiaetal98,char99}, though of limited
resolution, have found indications that this width may vary with
latitude and suggested that the tachocline could have a prolate
shape. We will show in section 4 how the change from a spherical
to a prolate shape of the tachocline and its thickness influence
the outputs of the model.

\subsection{Meridional circulation}

In the present stage of the observational techniques, there is a
set of independent measurements of the meridional circulation that
confirms a surface poleward flux of about $20$ m s$^{-1}$, which
persists until a depth of $0.85\sr$
\citep{hathaway96a,hathaway96b,latushko96,snodgrass96,
gilesetal97,kommetal93,braunfan98}. However, even helioseismology
has been unsatisfactory to measure a deeper flow. Mass
conservation predicts an equatorward return flow and a unique
convection cell per meridional quadrant. Here, we use the same
analytical prescription of \cite{nandy2002} and \cite{guerrero04},
which was previously introduced by \cite{dikchoud1}, and
\cite{choudhuri95}:

\begin{equation}
\rho(r) \bf{u}=\nabla \times [\psi(r,\theta)e_{\phi}]\label{eq4}\quad,
\end{equation}
where $\psi$ is a stream function given by:
\begin{eqnarray}
\psi r\sin\theta &=& (r-R_p)\psi_0
\sin\left[\frac{\pi(r-R_p)}{(\sr-R_p)}\right]\label{eq5} \\\nonumber
&\times&(1-e^{-\beta_1\theta^{\epsilon}})(1-e^{\beta_2(\theta-\pi/2)})
\\\nonumber
&\times&e^{[(r-ro)/\Gamma]^2} \quad,
\end{eqnarray}
and $\rho(r)$ is a density profile for an adiabatic sphere with a
specific heat ratio $\gamma$$=$$5/3$ (polytropic index $m$$=$$1.5$),
thus:
\begin{equation}
\rho(r)=C\left(\frac{\sr}{r}-0.95\right)^m\label{eq6} \quad .
\end{equation}

The value of $\psi_0$ and $C$ are chosen in such a way that the
amplitude of the meridional velocity, $u_{\theta}$ at middle latitudes
is $U_0$$=$$\psi_0/C$$=$$25$ m s$^{-1}$. The assumed value for the
other five parameters are: $\beta_1$$=$$6.06\times10^9$ cm$^{-1}$,
$\beta_2$$=$$4.6\times10^9$cm$^{-1}$, $\epsilon$$=$$2.0000001$,
$r_o$$=$$(\sr-R_{min})/4.0$, and $\Gamma$$=$$3.47\times 10^{10}$
cm. $R_{min}$ is the radius of the bottom of the convective
boundary,and $R_p$ is the radius of the penetration depth of the flow.

\begin{figure}
  \begin{centering}
  \includegraphics[width=6cm]{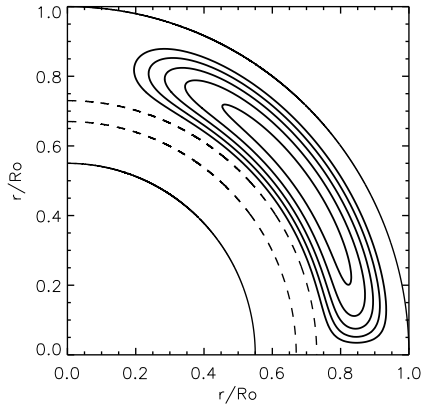}
  \caption{The adopted meridional circulation lines profile in this
  work. The dashed lines represent the tachocline with a constant
  thickness $d_1$$=$$0.05\sr$ in this case.}
  \label{fig2}
  \end{centering}
\end{figure}

In a previous work, it has been  shown that a deep meridional flow
solves only partially the problem of the distribution of the
toroidal magnetic field \citep{guerrero04}. Besides, as stressed
in \S1, this assumption seems to need physical support and does
not seem to be confirmed by numerical simulations either, so that,
in the present analysis we will not adopt a deep meridional
distribution (see however \cite{nandy2002}, and
\cite{chatterjeeetal04}, for an alternative interpretation).

The dynamical shallow water model of \cite{gilmanmiesh04} showed
that it is impossible for the meridional flow to go below a few
percent of the top for the solar tachocline. Recent calculations
of \cite{rudigeretal05} support this result arguing that a weak
penetration is able to produce a flux dominated solar dynamo
inside the solar convection zone, without the necessity of
participation of the tachocline, however they did not consider the
overshoot effect of plumes and jets that actually can penetrate
inside the tachocline and even down in a thin fraction of a more
sub-adiabatic medium  \citep{rogersetal06}. In our calculations,
we will assume a weak penetration inside the tachocline and, as
the width of the tachocline is allowed to vary in this work, the
percentage of penetration will be able to vary, as well, but we
will consider a constant value for the penetration radius
$R_p$$=$$0.69\sr$. Figure \ref{fig2} shows the assumed meridional
circulation profile.

\subsection{Magnetic diffusivity}

The radial dependence of the magnetic diffusivity is probably
the most undetermined of the profiles of the solar interior. Diffusion
must exist in the convective envelope due to the intense turbulence
present, but besides the values of super-granular diffusion
observed at the surface ($\eta_S \sim 10^{12}-10^{14}$ cm$^2$
s$^{-1}$), the depth dependence is uncertain and, in general, only
approximate values are assumed for this quantity in the
dynamo models. We will here follow \cite{dikpatietal04} and
assume a weak turbulent diffusivity regime for the radiative interior,
with $\eta_r$$=$$2.2\times 10^8$ cm$^2$ s$^{-1}$, a turbulent regime
for the convective envelope, with $\eta_{c}$$=$$5\times10^{9}$
cm$^2$ s$^{-1}$ above the tachocline, and a third region of
super-granular diffusivity, from $r_{c1}$$=$$0.95\sr$ on, where
$\eta_{S}$$=$$10^{12}$ cm$^2$ s$^{-1}$.

\begin{equation}
\eta(r)=\eta_r + \frac{\eta_{c}}{2}\left[1+erf
  \left(\frac{r-r_{c}}{d_2}\right)\right]
+\frac{\eta_{S}}{2}\left[1+erf
  \left(\frac{r-r_{c1}}{d_3}\right)\right]\quad .\label{eq7}
\end{equation}

\begin{figure}[ht]
  \begin{centering}
  \includegraphics[width=6cm]{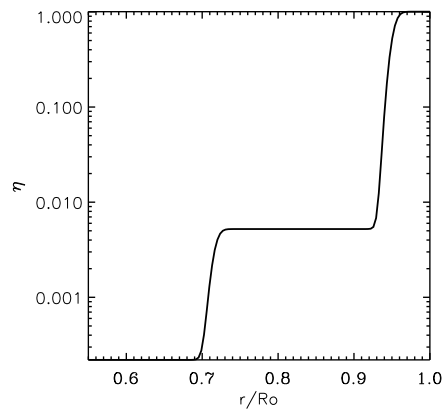}
  \caption{Adopted depth dependence of the magnetic diffusivity
  profile. Y axis is in log-scale and is normalized to the
  supergranular surface value ($10^{12}$ cm$^2$ s$^{-1}$).}
  \label{fig3}
  \end{centering}
\end{figure}

Different combinations of the parameters $d_2$, $d_3$, $r_c$, and
$r_{c1}$ in the equation above can be considered (see, e.g.,
 \cite{dikpatietal06}). The boundary between a weak and a strong
  turbulent regime is the location where  an abrupt change
  in the temperature gradient occurs in helioseismic calibrations.
We have presently considered {\bf $r_c$$=$$0.715 \sr$} and
$d_2$$=$$0.01\sr$ which are compatible with the values that have
been obtained from helioseismic measurements, the first with a very
high precision and the second as an upper limit  \citep{basu97}, and 
{\bf$r_{c1}$$=$$0.95\sr$}, $d_3$$=$$0.01\sr$
\citep{dikpatietal06} (see Figure \ref{fig3}). Our model is
specially sensitive to the the diffusivity variation between the
radiative core and the convective  envelope. With the choice of
the parameters above, we are considering a weak turbulent regime
for the lower region of
 the tachocline and a turbulent regime  for the upper part of it, with
 a sharp variation between them, in order to allow both storage of
 magnetic field and overshooting of material below the tachocline
(see, however, \S 4.3 for a more detailed analysis of the dependence
 of the model with the parameters $\eta_c$ and $r_c$).

\subsection{The source of poloidal field (the $\alpha$ term)}

From the Parker mean field dynamo theory days until now, the $\alpha$
source term of the poloidal field remains virtually undetermined. As a
coupling factor between eqs. (\ref{eq1}) and (\ref{eq2}), the main
function of this term
is to generate poloidal field from pre-existent emergent
magnetic flux tubes in the toroidal direction,
and its form is oriented to resemble the Babcock-Leighton concept
of formation of dipolar like field as the result of the decay of BMRs
and the reconnection of loops from both hemispheres.

The magnetic flux tubes, at the base of the convection zone are
amplified by the differential rotation until they reach a value such
that the magnetic pressure begins to dominate over the gravitational
pressure. Numerical simulations suggest a value of $10^5$ G. Then,
they are lifted on the entire convection zone and twisted by the
Coriolis force to form BMRs at the observed latitudes.

\begin{figure*}[ht]
  \begin{centering}
  \includegraphics[width=6cm]{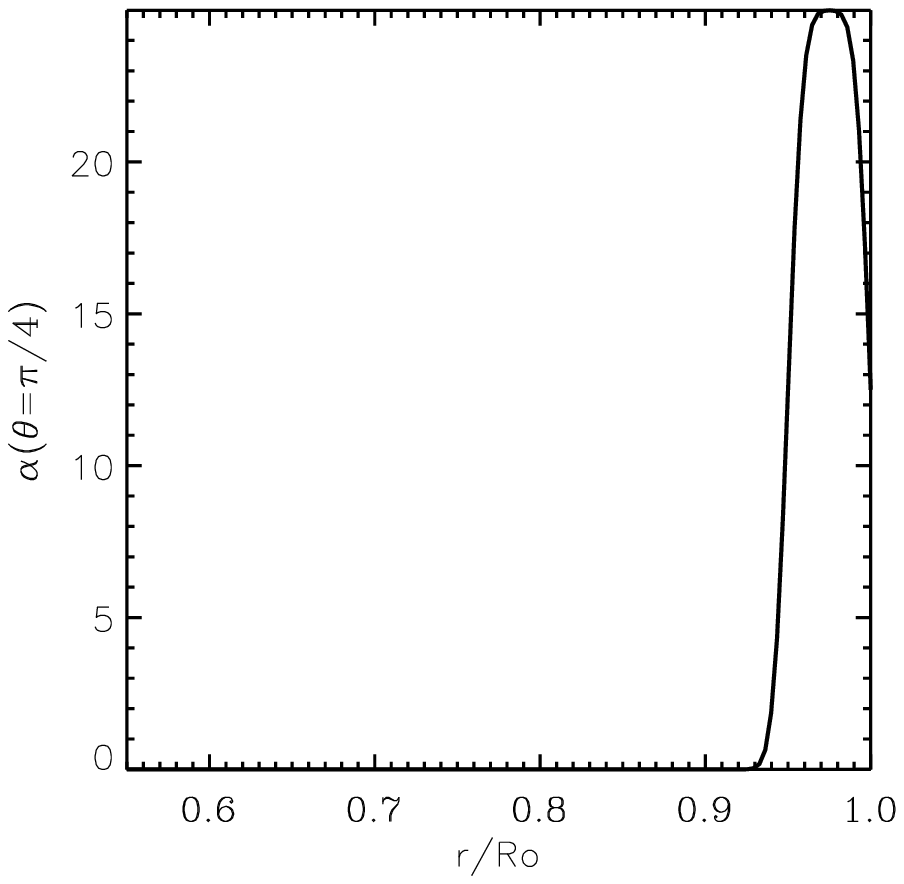}\hspace{2cm}
   \includegraphics[width=6cm]{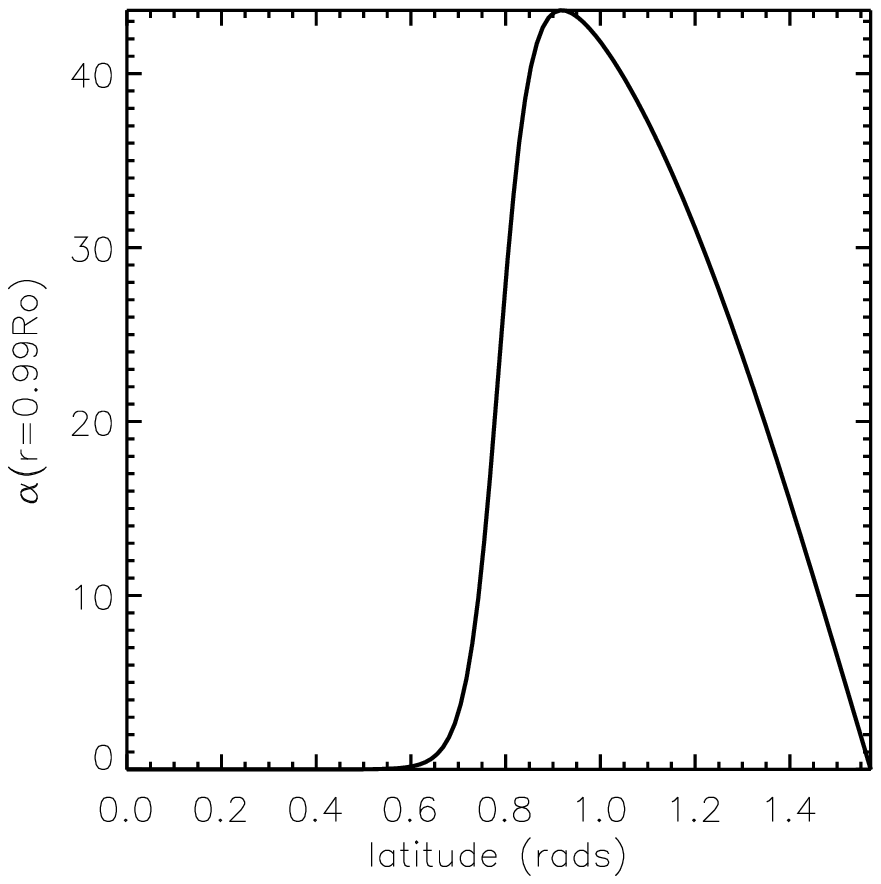}
  \caption{Assumed radial and latitudinal profiles for the $\alpha$
  effect mechanism. Radially, the $\alpha$ term is concentred close to
  the surface. Its latitudinal distribution corresponds to the belt
  where the sunspots appear.}
  \label{fig4}
  \end{centering}
\end{figure*}

In order to allow this process to occur, we let the poloidal field
at the surface ($r$$=$$\sr$) to be proportional to the toroidal
field at the base of the solar convection zone
($r$$=$$R_c+d_1/2$), where $d_1$ is the thickness of the
tachocline (see, e.g. \cite{dikchar99}). Since the toroidal field
lines are mainly concentrated in a belt around the solar equator,
following \cite{dikpatietal04}, we consider the action of the
Coriolis force also concentrated at lower latitudes, thus:
\begin{eqnarray}\label{eq8}
S_1(r,\theta,B)&=&\alpha_0
\frac{1}{4}\left[1+\mathrm{erf}\left(\frac{r-r_2}{d_2}\right)\right]
\left[1-\mathrm{erf}\left(\frac{r-r_3}{d_3}\right)\right] \\\nonumber
&\times& \sin\theta
\cos\theta\left[\frac{1}{1+e^{\gamma_1\left(\pi/4-\theta\right)}}\right]
\left(1+\left[\frac{B(R_c,\theta)}{B_0}\right]^2\right)^{-1} \quad,
\end{eqnarray}

\noindent where $r_2$$=$$0.95\sr$, $r_3$$=$$\sr$,
$d_2$$=$$d_3$$=$$0.01\sr$, $\gamma_1$$=$$30$. The amplitude of the
poloidal source is determined
by $\alpha_0$ for which we assume a fixed value of $130$ cm
s$^{-1}$. Figure \ref{fig4} depicts the radial and latitudinal
dependence of this profile.

The latitudinal dependence of $S_1$ is of great importance in the
reproduction of the features of the solar cycle. We tested
alternative possibilities reported in literature, i.e., profiles
which are proportional to $\sin\theta \cos\theta$
\citep{dikchar99}, $\sin^2\theta \cos\theta$ \citep{kuketal01},
and
 $\cos\theta$ \citep{chatterjeeetal04} factors, but we found that the
profile prescribed by eq. (\ref{eq8}) is the one that better
reproduces an observed butterfly diagram.

The last term on the RHS of eq. (\ref{eq8}) is a quenching term
limiting the growth of the poloidal field. Actually, this
quenching should be given by the back reaction of the magnetic
field on the velocity field, but the physical way by which the
poloidal field stops growing is unknown. This form, which was also
assumed previously by many authors, simply reproduces the fact
that a toroidal field with a value larger than $B_0$$=$$10^5$ G
would generate sunspot pairs with tilts which is in disagreement
with the joy's law \citep{dsilva93}. It is important to note that
this term is the only source of non-linearity of the system and
works non-locally.

Other models, like, e.g. \cite{nandy2002} and \cite{chatterjeeetal04}
employ a numerical procedure to include the quenching of the poloidal
field. In these models, whenever the toroidal
magnetic exceeds $B_0$ at the base of the solar convection zone, a
fraction $f$ of it is artificially made to erupt to the surface at time
intervals $\tau=8.8 \times 10^5$ s \citep{nandy2001}. In essence, this
buoyancy mechanism does the same as our quenching term above, mainly
with a supergranular diffusion value at the surface,  with a
time delay in order to ensure the right phase relation between the
toroidal and the poloidal fields, but this phase lag can be fitted by
tuning the velocity of the meridional flow at the base of the
convection zone.

We notice that in a recent work, \cite{dikpatietal04} have
introduced an $\alpha$ effect located in the tachocline. It could
occur as a consequence of hydrodynamic instabilities at the top of
the tacholine and become important in two different ways. First,
as this term does not depend on the toroidal field magnitude, it
could be a source of field in times when activity goes down (i.e.
during the Mounder minimum), second, it could provide the correct
parity relation (Hale's law) when the integration of the equations
is made in the two hemispheres. Both of  these issues are out of
the scope of this work and, for this reason have not been
considered.


\section{Results}

\begin{figure*}
  \begin{centering}
  \includegraphics[width=10cm]{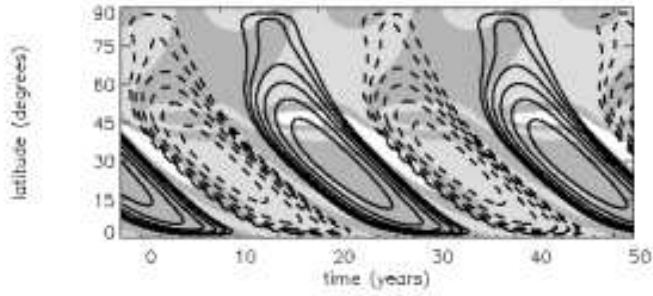}
  \caption{Time-latitude butterfly diagram for the model of Table
    1 with a tachocline of constant width $d_1$$=$$0.05\sr$. The
    continuous (dashed) lines represent the positive (negative)
    strength of the toroidal field at the base of the solar convection
    zone (i.e. the top of the tachocline $r$$=$$R_c+d_1/2$).  The
    lines are log spaced and cover the interval between $5 \times
    10^4$ $-$  $10^5$ G. The background gray scale
    represents the positive (clear) and negative (dark) radial field
    at the surface.}
  \label{fig5}
  \end{centering}
\end{figure*}

Employing the profiles described above, we have first built the
butterfly diagram of the model  depicted in Figure \ref{fig5} and
Table 1, with a constant width tachocline with $d_1 = 0.05
R_{\odot}$. It reproduces some of the main features of the large
scale solar dynamo model. That is, the periodicity of the magnetic
cycle, the observed magnitude of the toroidal fields near the
equator, as well as, the weak radial fields near  the pole, and
the phase lag between them. However, we notice that large toroidal
fields persist above 45 degrees suggesting that sunspots should
also appear  at those latitudes, which is not observed. In the
next section we explore alternative possibilities to this
solution.

\begin{table}
\centering
  \caption{Values of the parameters used in the model of
  Figure 5.}
  \label{table1}
  \begin{tabular}{lc}\hline
       PARAMETER     &   VALUE \\\hline
       $\Omega_{Eq}$ & $2\pi \times 460.7$ nHz \\
       $R_c$         & $0.7\sr$ \\
       $d_1$         & $0.05 \sr$ \\\hline
       $U_0$         & $25$ m s$^{-1}$ \\
       $R_p$         & $0.69 \sr$ \\\hline
       $\eta_r$      & $2.2 \times 10^8$ cm$^2$ s$^{-1}$  \\
       $\eta_c$      & $5.0 \times 10^{9}$ cm$^2$ s$^{-1}$ \\
       $\eta_s$      & $1.0 \times 10^{12}$ cm$^2$ s$^{-1}$ \\\hline
       $\alpha_0$    & $130$ cm s$^{-1}$ \\\hline
  \end{tabular}
\end{table}

\subsection{An ellipsoidal tachocline}

As stressed in \S1, so far, the exact location and width of the
tachocline are still unknown. A good revision of the observational
results and theoretical models of the tachocline can be found in
\cite{cobardetal01}. Some evidence for a prolate shape has been
found by \cite{antiaetal98} and \cite{char99}, but this could be
due to observational uncertainties or to a particular sensitivity
to the inversion techniques employed to analyze these data
\citep{cobardetal01}. Numerical simulations \citep{dikgilman01}
have found that a strong magnetic stress ($B>10^5$ G) can pile the
matter into the pole increasing the density and thus the width of
the tachocline there. This could explain, in principle, the
prolate shape, however, since a strong field is present only
during the maximum of activity, this could suggest that the
tachocline would have a varying shape through the cycle, becoming
nearly prolate during the maximum activity.

In order to evaluate the effects of a potential prolate tachocline
in a flux dominated dynamo model, we have introduced a latitudinal
dependence on the width of the tachocline in equation (\ref{eq7}),
making it to vary from $d_1$(pole)$=$$0.07\sr$ to
$d_1$(equator)$=$$0.02\sr$ (see left panel of Figure \ref{fig6},
top). This modification in the shape of the tachocline has the
effect of decreasing the radial shear $\rsh$ at the high
latitudes. If the appearance of strong toroidal fields at high
latitudes were mainly sensitive to the radial shear, then one
should expect that a reduction in $\rsh$ at those latitudes would
reduce the amplification of the toroidal field. However the
butterfly diagram depicted in Figure \ref{fig6} (top), which was
calculated for a prolate tachocline, shows no significant changes
with respect to the diagram of Figure \ref{fig5}, for which a
constant width tachocline was assumed. Besides, when an oblate
configuration is considered instead (left panel of Figure
\ref{fig6}, bottom), we find that the toroidal field is amplified
to its maximum value only below the $60^{\circ}$ latitude (right
panel of Figure \ref{fig6}, bottom). In other words, for an oblate
tachocline, for which the radial shear is improved towards the
higher latitudes, we find an inhibition in the generation of
toroidal fields at those latitudes, contrary to what is commonly
expected. Also, for both the oblate and the prolate tachoclines,
the generation of toroidal fields at the low latitudes is
practically the same, suggesting that $\rsh$ is not influencing
the behavior of the toroidal field.

\subsection{A thinner or a thicker tachocline?}

\begin{figure*}[htb!]
\begin{center}
\includegraphics[width=3.5cm]{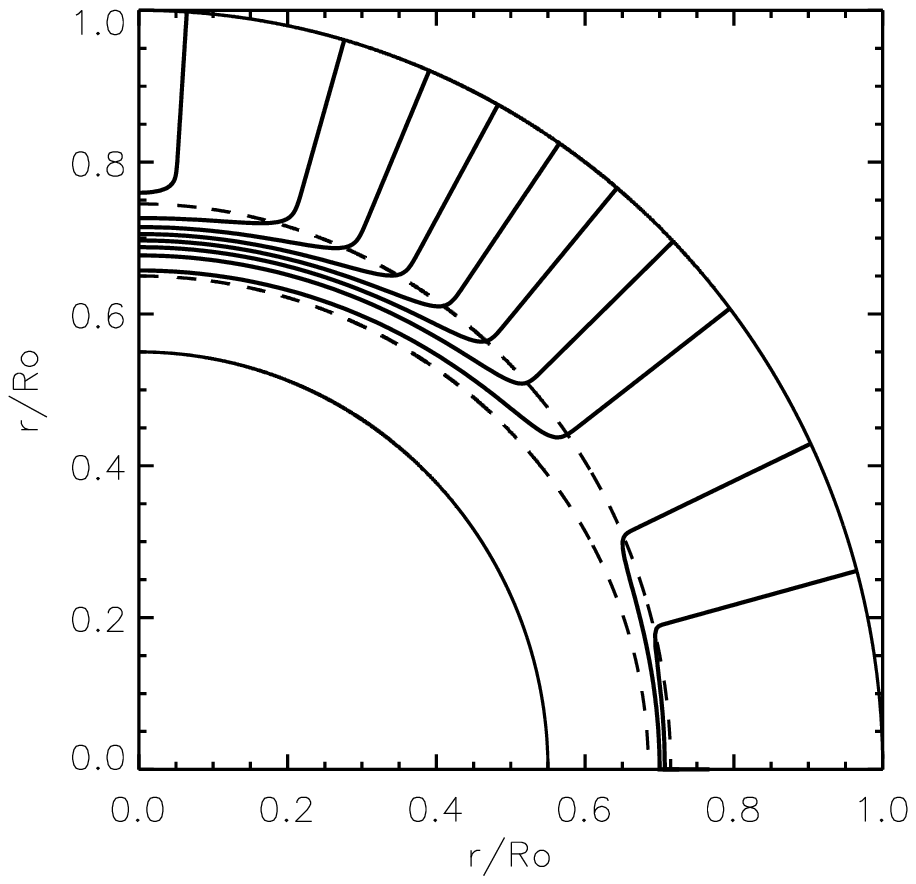}\vspace{0.3cm}
\includegraphics[width=9.0cm,height=3.5cm]{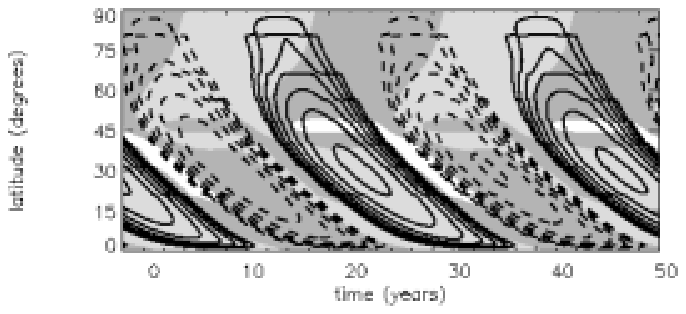}\\
\includegraphics[width=3.5cm]{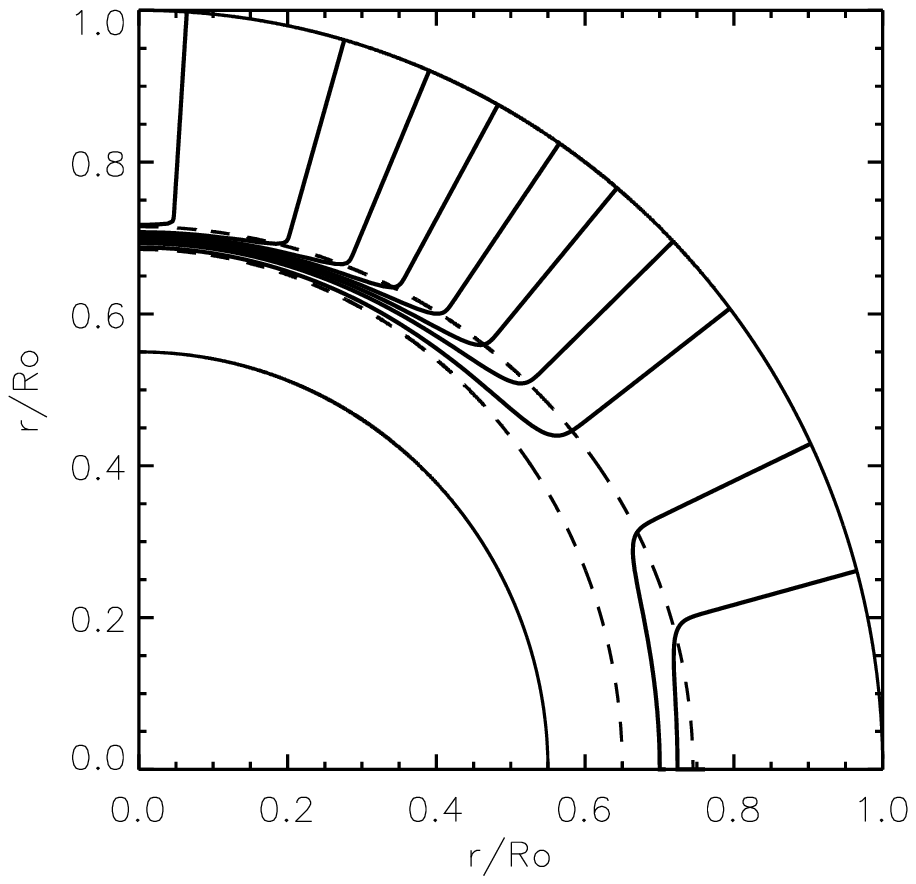}\vspace{0.3cm}
\includegraphics[width=9.0cm,height=3.5cm]{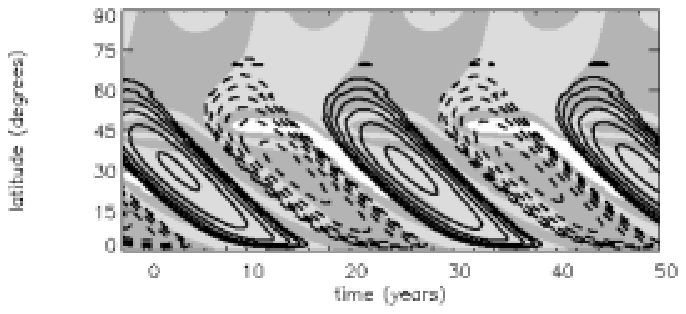}\\
\caption{Angular velocity profiles and time-latitude butterfly diagrams
  for a prolate tachocline (top panel); and an oblate tachocline,
  (bottom panel). See the text for details. The contours
  specifications in the right panels are the same as in Figure
  \ref{fig5}. The tachocline in the left panels is represented by
  dashed lines.}
\label{fig6}
\end{center}
\end{figure*}

Two interesting remarks can be pointed out from the results of
Figure \ref{fig6}. On one hand, it does not seem to be posssible
to reproduce the butterfly diagram with a prolate tachocline
(i.e., with a thicker tachocline at higher latitudes), at least
not for the assumed conditions. On the other hand, we find that
 the radial shear
$\rsh$ does not seem to be contributing for the amplification of
the toroidal magnetic field.

In order to investigate the origin of this apparent paradox, we
have considered again a spherical tachocline (with constant
width), and then computed the components of the shear term on the
RHS of equation (\ref{eq2}). $({\bf B_p} \cdot \nabla)
\Omega$$=$$B_r\rsh + B_{\theta}/r \, \partial \Omega / \partial
\theta$, at a high latitude from
the equator ($\theta$$=$$60^{\circ})$ \footnote{This is the
latitude above which strong toroidal fields $\gtrsim 5 \times 10^4$ G
should not appear, but have developed in the butterfly diagram of
Figure \ref{fig6} (top).},
for $r$$=$$R_c$ (i.e., in the center of the tachocline where the
radial shear is maximum) at a time when the radial magnetic field
($B_r$) inverts its polarity. At this time of the cycle, the action
of the shear term upon the poloidal field generates new branches of
toroidal
field and establishes the morphology of this growing branch for the
next phase of the cycle. We have
also computed the toroidal magnetic field ($B$), at the same latitude
when it reaches its maximum at the top of the
tachocline. These
quantities were computed as a function of the width of the
tachocline ($d_1$) taken in the range of
possible values that are inferred from the observations ($0.01\sr
\le d_1 \le 0.1 \sr$; \cite{cobardetal01}), the results are
depicted in Figure \ref{fig7}.

We note that the radial component, $B_r (\frac{\partial
  \Omega}{\partial r})$ (Figure \ref{fig7}$a$), actually tends to
  decrease with the increase of the width, as
  expected. But, more significant is the fact that its value is about
  two orders of magnitude smaller than that of the latitudinal
  component $\frac{B_{\theta}}{r} (\frac{\partial \Omega}{\partial
  \theta})$ (Figure \ref{fig7}$b$). Therefore, when a new toroidal
  field begins to be generated its growth is dominated by the
  latitudinal shear in eq. (\ref{eq2}). Tests run for different
  latitudes revealed the same effect. This is confirmed by Figure
  7$c$, which shows that the magnitude of the generated toroidal field
  at high latitudes varies with the width of the tachocline in a
  similar way to the latitudinal shear component (Figure 7$b$),
  attaining a maximum value at a width, $d_1$ that depends on the
  assumed magnetic diffusivity. This latter result will be explored in
  more detail in the next paragraphs.

The results above raise two new questions. Since the radial shear
does not seem to be important for the amplification of the
toroidal magnetic field, nor even at high latitudes, is the
tachocline really participating in the dynamo process? And if yes,
what is the real thickness of this layer? In the present model,
the tachocline is not only the interface where the radial shear
$\rsh$ is maximized, but also the place where the meridional flow
penetrates and  the magnetic flux tubes are stored and amplified
before erupting by buoyancy effects to the upper layers, so that
our answer to the first question is yes. In order to answer to the
second question we can examine the amplitudes of the generated
toroidal field at the latitude $60^{\circ}$, at the top of the
tachocline in Figure \ref{fig7}$c$. It indicates that
values of $d_1$$\lesssim$$0.02$$\sr$ and $d_1$$\gtrsim$$0.08 \sr$
\footnote{As it will be shown below, this result holds only for
  values of $\eta_c$$\lesssim$$2 \times 10^{10}$ cm$^2$ s$^{-1}$.}
  are appropriate to prevent the formation of strong toroidal fields
  at high latitudes, as required by the observations. But which value
  of $d_1$ is the best one? As we have mentioned above, the toroidal
  fields that develop at the high latitudes are also dependent of the
  adopted diffusivity profile. For a definitive answer, we have to
  wait for better helioseismic observations in the future,
  nonetheless, a more detailed scanning of the diffusivity parameters,
  as shown in the next section, can shed some light on this question.

\subsection{Parameter dependence}

\begin{figure}[htb!]
\begin{center}
\includegraphics[width=9.0cm,height=3.5cm]{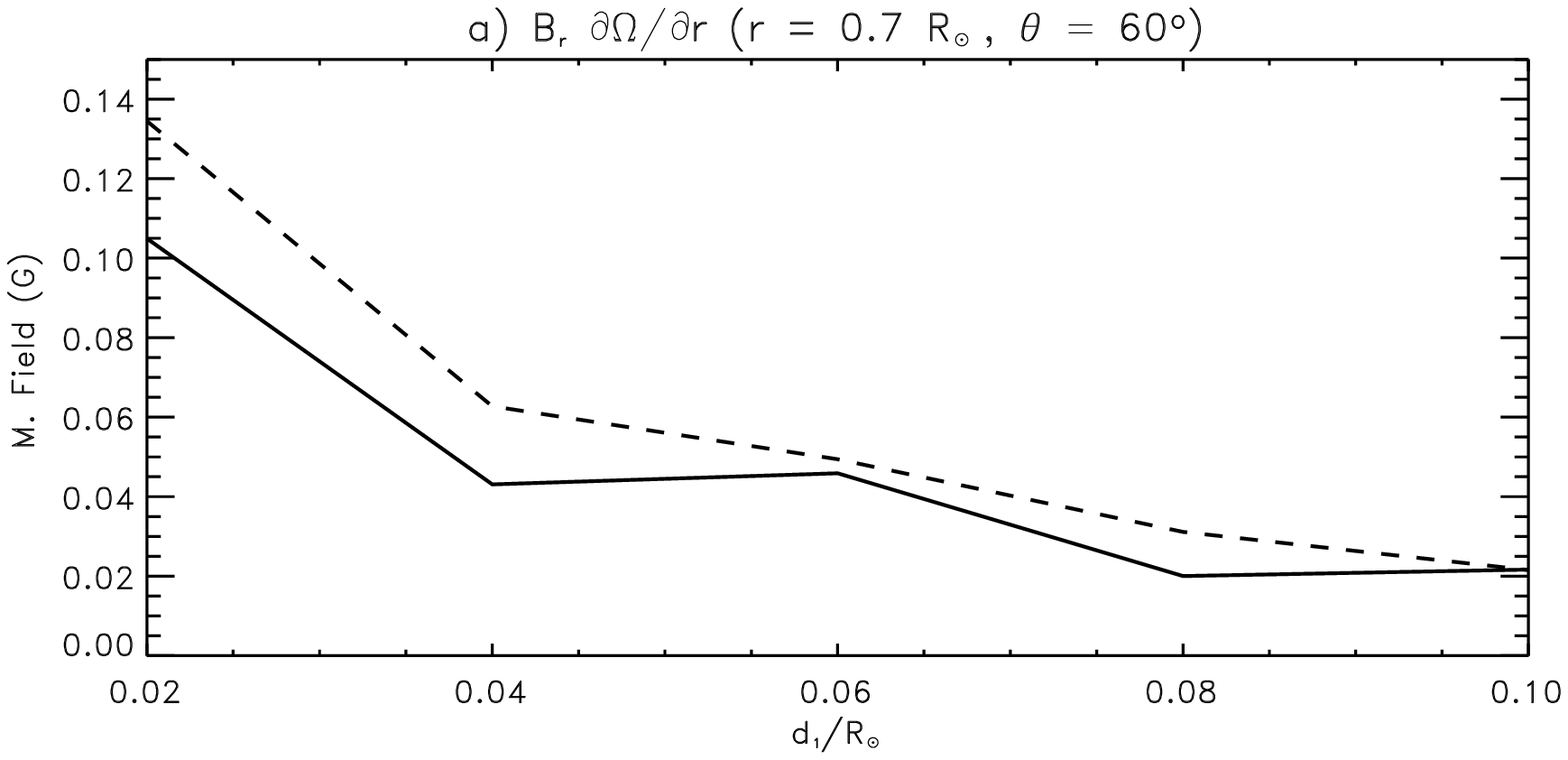}\\
\includegraphics[width=9.0cm,height=3.5cm]{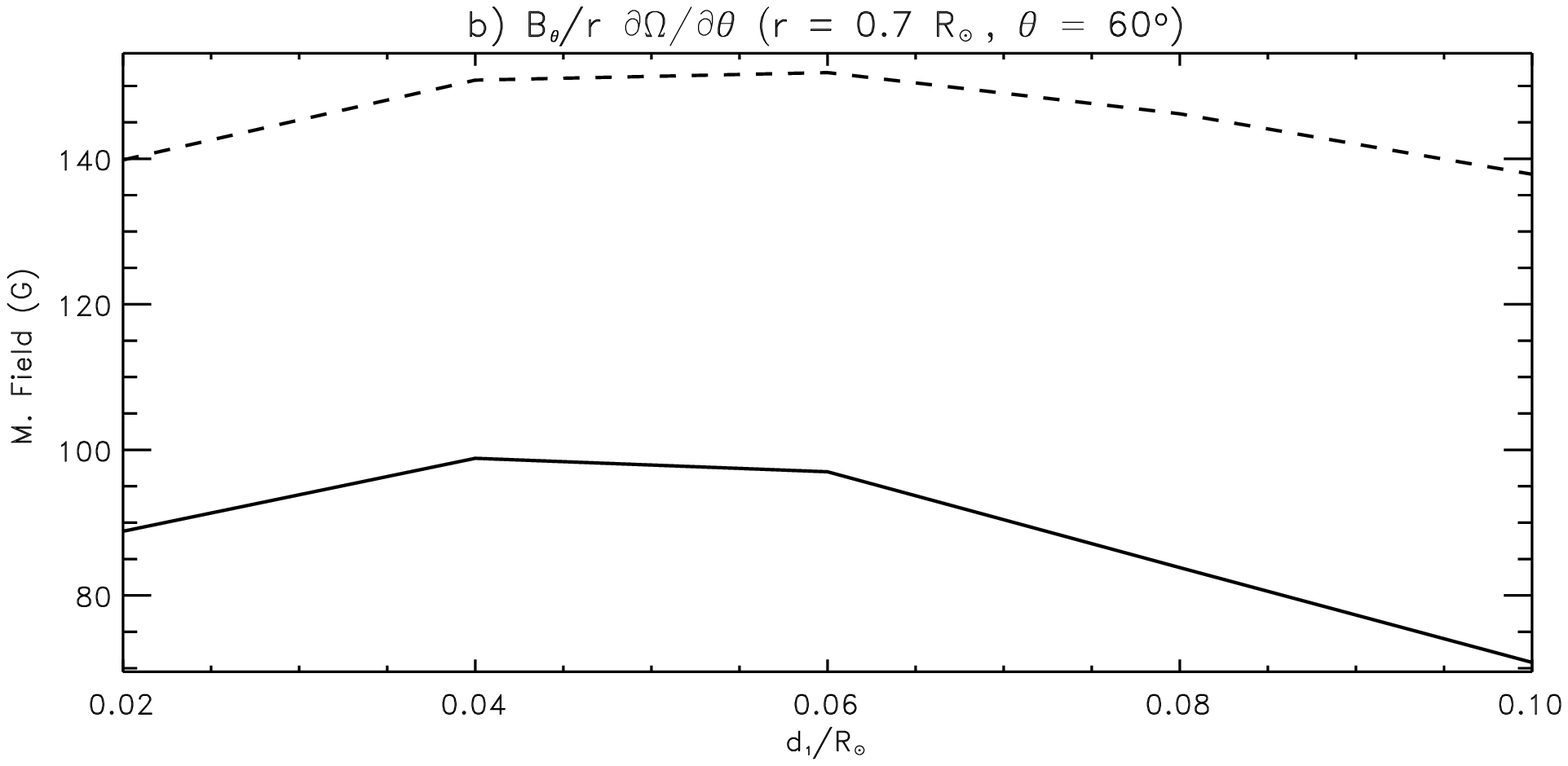}\\
\includegraphics[width=9.0cm,height=3.5cm]{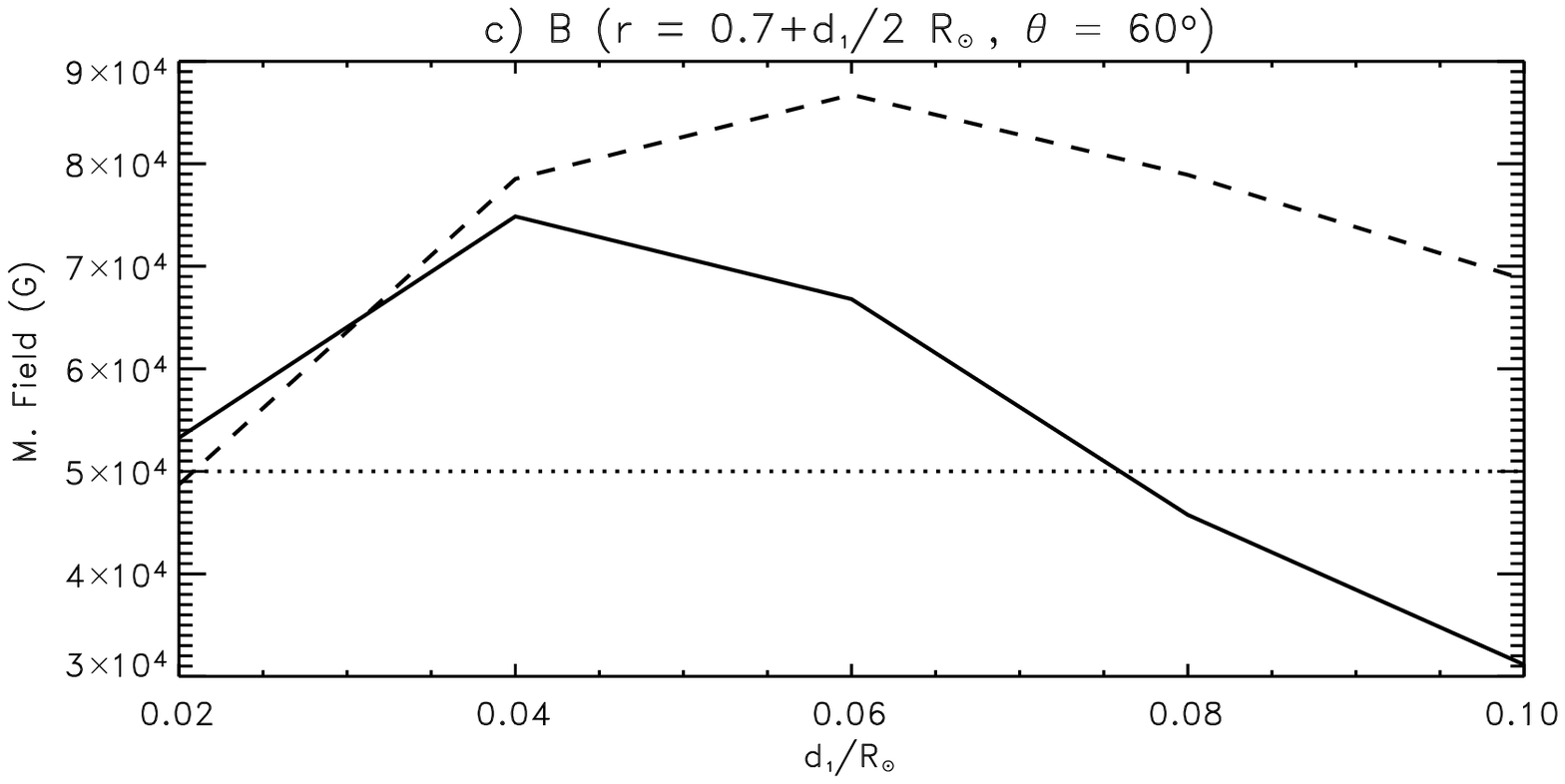}\\
\caption{Amplitude of the shear terms: a) $B_r
\left(\frac{\partial \Omega}{\partial r}\right)$;  b)
$\frac{B_{\theta}}{r} \left(\frac{\partial \Omega}{\partial
\theta}\right)$ at the time of inversion of the radial magnetic field,
at a high latitude ($60^{\circ}$),  at the center of the tachocline
  and c) the toroidal field $B$ at the time it reaches its maximum
  at the top of the tachocline as a function of the thickness of the
  tachocline. Dashed and continuous lines represent two different
  values of the turbulent diffusivity in the convection zone,
  $\eta_c$$=$$5\times 10^9$ cm$^2$ s$^{-1}$, and $2\times 10^{10}$
  cm$^2$ s$^{-1}$ respectively (the continuous line corresponds to the
  value of Table \ref{table1}).}
\label{fig7}
\end{center}
\end{figure}

The present model employed a large number of parameters and we have
performed some tests in order to explore how sensitive the
results above are to them.
The meridional circulation term in equations (\ref{eq4})-(\ref{eq6})
has two free parameters, the amplitude of the meridional
flow at the surface, at middle latitudes, $U_0$, and the depth of the
penetration of the meridional flow, $R_p$. While the first does not
affect the strength of the generated magnetic fields (though it is very
important to establish the period of the cycle), the second
is able to change the latitude of formation of strong magnetic fields,
as stressed by \cite{nandy2002}. We should note, however, that in this
analysis we are assuming a weak penetration regime, and small
variations around the value depicted in Table \ref{table1} do not
change significantly the results above.

The $\alpha$ term, whose semi-empirical profile
was chosen in order to better reproduce the observations (Figure
\ref{fig4}), has only one free parameter, the amplitude, $\alpha_0$
(in equation \ref{eq8}). It determines the exact amount of poloidal
field to be regenerated in order to support the cycle. We find that
our model is generally very insensitive to variations to it.

For the diffusive terms of equation (\ref{eq7}), it is possible to
establish some constraints in the different regimes. For the most
external layer, we have adopted a value for the diffusivity that
has been obtained from the observations, so that it has been fixed
in all the simulations. The value of the diffusivity at the
radiative zone is not expected to affect the results either, but
these can be sensitive to the assumed turbulent diffusivity at the
convection zone ($\eta_c$), as indicated by Figure \ref{fig7}. We
may choose an appropriate value for $\eta_c$ based on the period
of the cycle and the magnitude of the generated fields. If
the value of $\eta_c$ is too large, the system will enter in a
diffusion dominated regime, thus reducing the period of the cycle
in the butterfly diagrams. On the other hand, if $\eta_c$ is too
small, it will lead to too large values of the radial field at the
poles.

\begin{figure}[htb!]
\begin{center}
\includegraphics[scale=0.7]{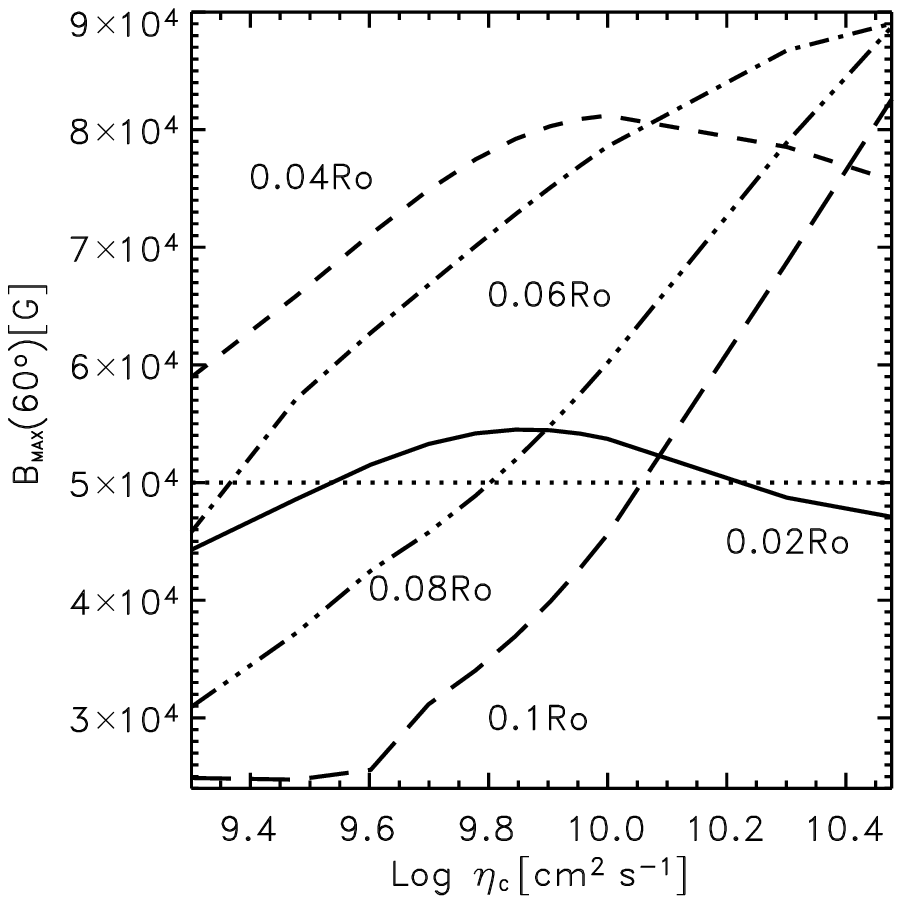}\vspace{0.3cm}\\
\includegraphics[scale=0.7]{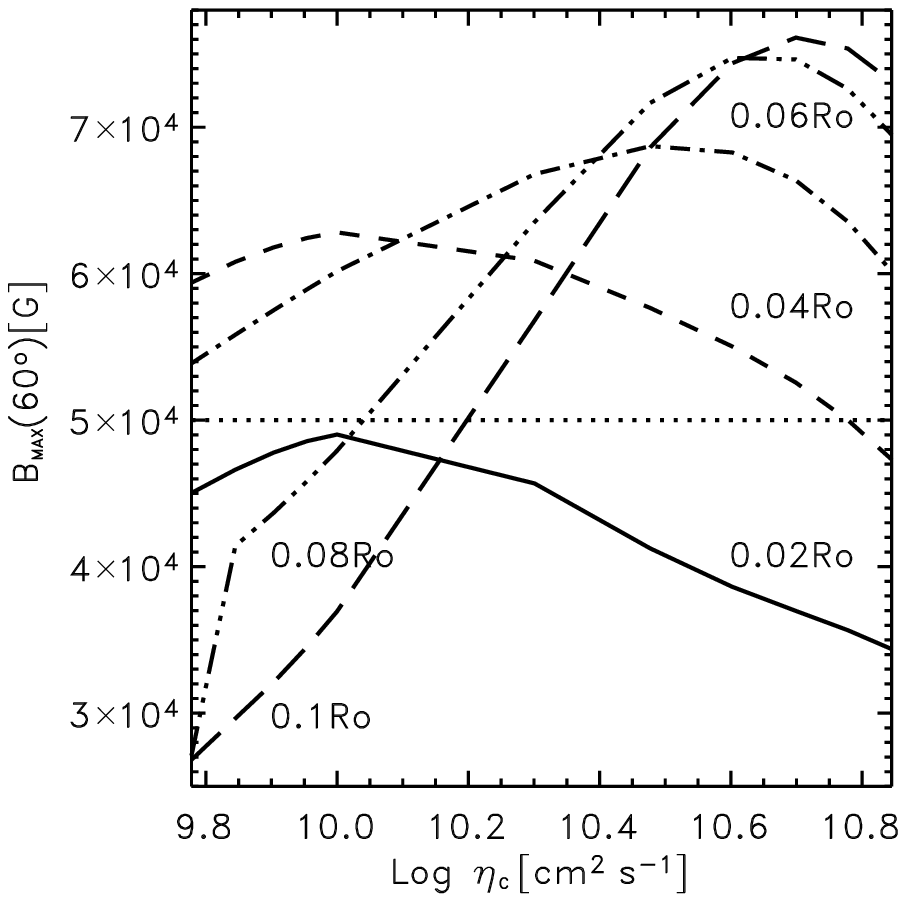}\vspace{0.3cm}
\caption{Maximum of the toroidal magnetic field at the top of the
tacholine as function of the diffusivity (in log-scale)at a latitude
of $60^{\circ}$, The different line styles represent different
widths of the tachocline $d_1$. The doted line represents the limit
between buoyant and no-buoyant magnetic fields $5\times10^4$ G, as
explained in the text. Only the values below this line will appear in
the desired latitudes. The top and bottom panels correspond to two
different values of $r_c$ in the diffusivity profile,
$r_c$$=$$0.715\sr$ (as in all previous Figures) and $r_c$$=$$0.72\sr$,
respectively.}
\label{fig8}
\end{center}
\end{figure}

In Figure \ref{fig8}, we have plotted the maximum toroidal
magnetic field at the top of the tachocline, at a latitude of
$60^{\circ}$ (as in Figure \ref{fig7}$c$), as a function of the
diffusivity $\eta_c$, taken in the range of values which are
appropriate for the solar cycle. Different line styles correspond
to different values of the tachocline width. The dotted line at
$5\times10^4$ G marks the limit between buoyant and no-buoyant
magnetic flux tubes. If a curve is located above this limit,
magnetic contours will appear above $60^{\circ}$ in the butterfly
diagram, and this is not desired. The top and bottom panels of
Figure \ref{fig8} have been plotted for two different values of
the parameter $r_c$ of the turbulent diffusivity profile of the
convection zone (eq. \ref{eq7}). According to Figure \ref{fig3},
this is the transition radius between a weak and a strong
turbulent regime. Though this value has been determined from
helioseismic observations with high precision ($r_c$$=$$0.715\sr$), we
have decided to vary it by $0.5$\% in order to check the sensitivity of
the results to it. In the top panel, $r_c$$=$$0.715\sr$ is the same 
value adopted in the previous Figures. In the bottom panel, we
have displaced this value to $r_c$$=$$0.72\sr$ which means that a
larger portion of the tachocline must lie in the less turbulent
(sub-adiabatic) zone. Both, the top and the bottom panels indicate
a similar behaviour for the different values of the tachocline
width, but with a slight shift of the curves of the bottom panel
towards larger diffusivity values, which are allowed only if the
turbulent zone is smaller.

The top panel shows that a tachocline with a width of about $2$\%
of the solar radius or less will produce solar like butterfly
diagrams for almost the entire diffusivity range. Intermediate
widths, $d_1$$\simeq$$0.04\sr$$-$$0.06\sr$, are out of the allowed
range of magnetic fields for any diffusivity, and larger widths
between $\sim$$0.08\sr$ to $\sim$$0.1\sr$, will also produce
butterfly diagrams which are in good agreement with the
observations for $\eta_c$ from $2\times10^9$ cm s$^{-2}$ to
$1\times10^{10}$ cm s$^{-2}$. The bottom panel also indicates that
intermediate widths ($d_1$$\simeq$$0.04\sr$$-$$0.06\sr$) will
produce inappropriate butterfly diagrams, while a thin enough
tachocline ($d_1$$\lesssim$$0.02\sr$) will produce solar like
butterfly diagrams for the entire range of appropriate
diffusivities, and thicker tachoclines
$d_1$$\simeq$$0.08\sr$$-$$0.1\sr$ will produce solar like results
only for diffusivities in the $6\times10^9$$-$$1.6\times10^{10}$
cm$^2$ s$^{-1}$ interval. Computations made with $r_c$$=$$0.71\sr$
revealed a similar behaviour with a slight shift of the curves
towards lower values of the diffusivity with respect to the
$r_c$$=$$0.715$ panel  (Figure \ref{fig8}, top).

The butterfly diagrams plotted in Figure \ref{fig9} for two
spherical tachoclines, one with a thin width ($d_1$$=$$0.02\sr$,
$\eta_c$$=$$5\times 10^9$ cm$^2$ s$^{-1}$, $r_c$$=$$0.72\sr$) and
the other with a thick width ($d_1$$=$$0.08\sr$,
$\eta_c$$=$$3\times 10^9$ cm$^2$ s$^{-1}$, $r_c$$=$$0.715\sr$) are
the ones that better reproduce the observations.

We notice that the results above are also applicable to
prolate and oblate tachoclines and naturally explain the results
of Figure \ref{fig6}. In fact, since a prolate tachocline has a
larger width at higher latitudes (top panel of Figure \ref{fig6}),
then the toroidal field contours with strength between $5 \times
10^4$ G and $1 \times 10^5$ G develop over the entire hemisphere
because the latitudinal shear (which dominates over the radial
shear in all latitudes) increases towards the poles. On the other
hand, in the oblate case (bottom panel of Figure \ref{fig6}) where
the width is smaller at higher latitudes, the toroidal fields are
suppressed there due to the smaller latitudinal shear, thus
resulting in a more concentrated field distribution at the lower
latitudes. However, the thickness of the tachocline at which the
latitudinal shear has a maximum or minimum value depends on the
assumed magnetic diffusivity profile. For the one assumed in
Figure \ref{fig6}, a minimum value is obtained for a thinner
tachocline (as in Figure \ref{fig9}, top) and therefore, the
oblate configuration is the one that better reproduces the
observations, but this scenario could change if a different
diffusivity value had been adapted for the tachocline zone, than
the one assumed in Figure \ref{fig6}. Note that the parameters
used to built the butterfly diagram of the prolate configuration
($d_1$$=$$0.07\sr$ at the poles, $\eta_c$$=$$5 \times 10^9$ cm$^2$
s$^{-1}$ ) lie in the forbidden zone of Figure \ref{fig8}. For
example, according to the top panel of Figure \ref{fig8}, if we
had taken $\eta_c$ between $2 \times 10^9$ cm$^2$ s$^{-1}$ and $1
\times 10^{10}$ cm$^2$ s$^{-1}$, then a prolate configuration with
a tachocline width at the poles with $d_1$ between $(0.08-0.1)\sr$
would produce an appropriate butterfly diagram, because in this
case even at the high latitudes, where the tachocline is thicker,
the latitudinal shear would be small enough to suppress the
toroidal fields there (like in Figure \ref{fig9}, bottom).


\section{Conclusions}

In this work we have explored the effects of variations in both
the shape and the width of the solar tachocline in a
flux-dominated kinematic solar dynamo model. First, employing an
improved version of the numerical approach of \cite{guerrero04}
with a choice of more realistic diffusion and $\alpha-$effect
profiles and assuming a tachocline with constant width
$d_1$$=$$0.05\sr$, we were able to reproduce successfully some of
the main features of the 11-year large scale solar magnetic cycle,
like the phase lag between the toroidal and the poloidal fields,
the correct period  and magnetic field magnitudes (Figure
\ref{fig5}). However remains of toroidal field component persisted
at the high latitudes of the butterfly diagram for these initial
conditions.

\begin{figure*}[htb!]
\begin{center}
\includegraphics[width=3.5cm]{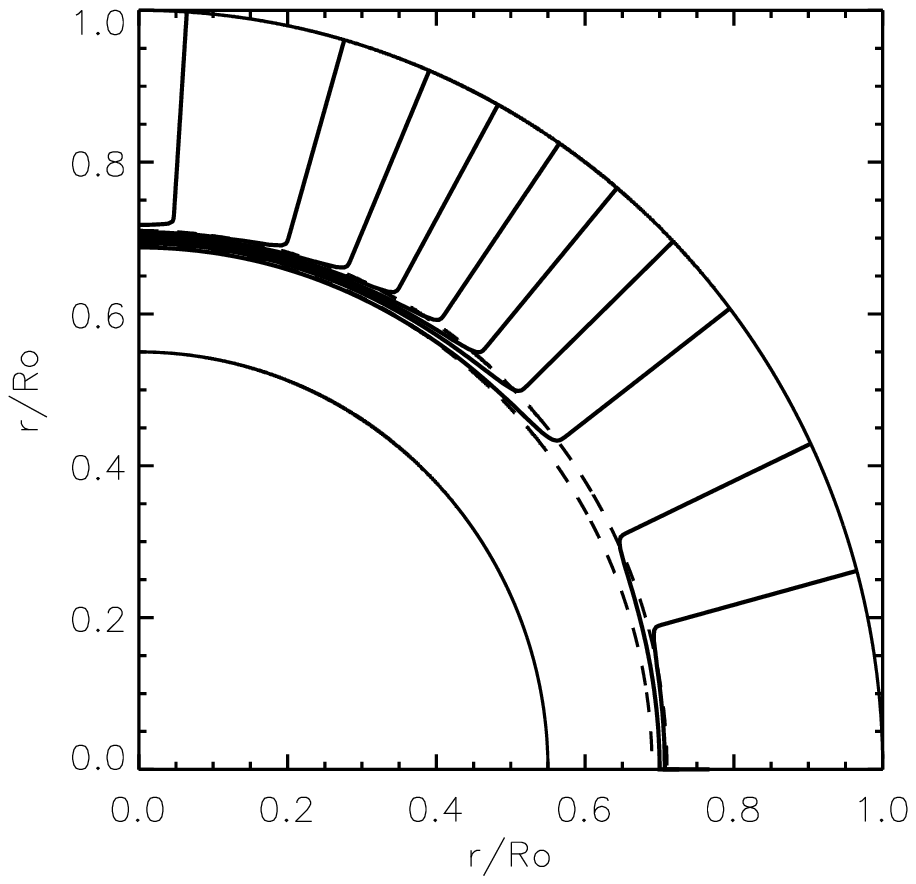}\vspace{0.3cm}
\includegraphics[width=9.0cm,height=3.5cm]{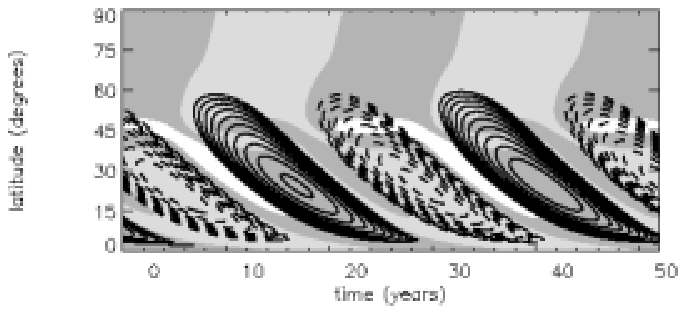}\\
\includegraphics[width=3.5cm]{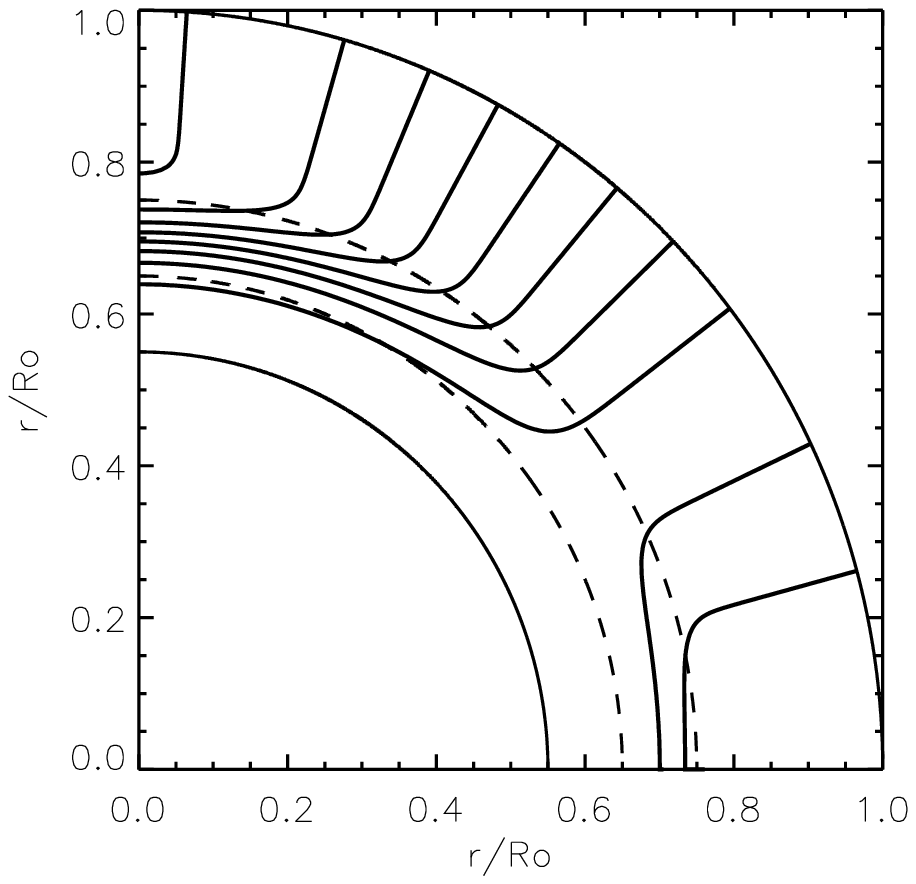}\vspace{0.3cm}
\includegraphics[width=9.0cm,height=3.5cm]{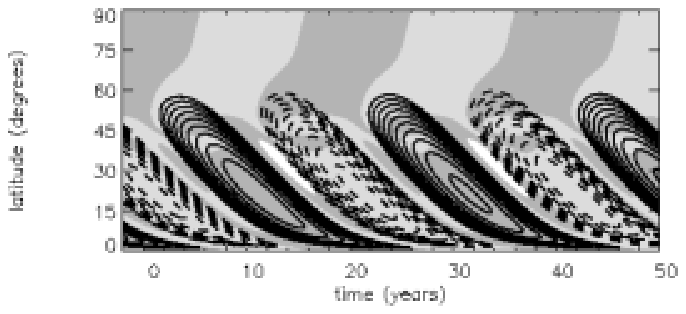}\\
\caption{Angular velocity profiles and time-latitude butterfly diagrams
  for a thin ($d_1$$=$$0.02\sr$, $\eta_c$$=$$5\times 10^9$ cm$^2$
  s$^{-1}$, $r_c$$=$$0.72\sr$) tachocline (top panel); and a thick
  ($d_1$$=$$0.08\sr$, $\eta_c$$=$$3\times 10^9$ cm$^2$ s$^{-1}$,
  $r_c$$=$$0.715\sr$)  tachocline, (bottom panel). See the text for
  details. The contours specifications in the right panels are the
  same as in Figure \ref{fig5}. The tachocline in the left panels is
  represented by dashed lines.}
\label{fig9}
\end{center}
\end{figure*}

Then, considering a prolate tachocline (Figure \ref{fig6}, top),
with a larger width in the polar region, which implies a smaller
radial shear $\rsh$ at these latitudes, we have obtained a
butterfly diagram with similar distribution of the toroidal fields
to that obtained with the constant width tachocline of Figure
\ref{fig5}. On the other hand, when considering an oblate
tachocline (Figure \ref{fig6}, bottom), we obtained a toroidal
magnetic field with a latitudinal distribution that is in better
agreement with the observations,  with an absence of toroidal
fields at latitudes higher than $60^{\circ}$.

In view of these surprising results, we computed the toroidal
field and the radial and latitudinal components of the shear term
(eq. \ref{eq2}) that is responsible for the amplification of the
toroidal field, $({\bf B_p} \cdot \nabla) \Omega$$=$$B_r\rsh
+ B_{\theta}/r \, \partial \Omega / \partial \theta$ as a
function of the width ($d_1$) of the tachocline at a high
latitude, at a position where $\rsh$ should be maximum.
The results suggest that the latitudinal component of the shear
term dominates over the radial term for producing toroidal field
amplification.

We have also found that these results are very sensitive to
the adopted diffusivity profile, specially in the inner convection
zone (which is characterized by the diffusivity $\eta_c$ and the
radius $r_c$ of transition between a weak and a strong turbulent
region). A diagram of the toroidal field at a latitude of
$60^{\circ}$ versus the diffusivity at the convection layer for
different values of the tachocline width has revealed that these
fields are mainly eliminated for tachoclines with width
$d_1$$\gtrsim$$0.08\sr$ and
$\eta_c$$=$$2\times10^9$$-$$1\times10^{10}$ cm$^2$ s$^{-1}$, for
$r_c$$=$$0.715\sr$; and $\eta_c$$=$$6\times10^9$$-$$1.6\times10^{10}$
cm$^2$ s$^{-1}$ for 
$r_c$$=$$0.72\sr$; or for $d_1$$\lesssim$$0.02\sr$ and practically any
value of $\eta_c$ in the appropriate solar range. For intermediate 
values of $d_1$$\simeq$$0.04-0.06\sr$, strong toroidal fields
should survive at high latitudes in the butterfly diagram and
those values are therefore not suitable. The best fits to the
observed butterfly diagram are shown in Figure \ref{fig9} for a
thin and a thick tachocline. We also conclude that a prolate
tachocline can produce solar like results depending on the choice
of the diffusivity profile and the adopted range of the tachocline
width.

Finally, we should note that the poloidal magnetic field
magnitudes are correctly reproduced except for a branch of strong
radial fields migrating to the equator (see white features in the
background grey scale of the butterfly diagrams of the figures).
The formation of this branch is related to our non-local
implementation of the $\alpha$-term. We will explore alternative
forms for this term in future work.

\begin{acknowledgements}
We would like to thank M. Dikpati for her useful comments on a former
version of this paper. We are also indebted to an anonymous referee
for his/her important and valuable comments. This work was partially
supported by CNPq grants.
\end{acknowledgements}

\bibliographystyle{aa}
\bibliography{5834r3}

\end{document}